\documentclass[letterpaper]{article}
\usepackage{aaai}
\usepackage{times}
\usepackage{helvet}
\usepackage{courier}

\usepackage{graphicx}
\usepackage{url}

\begin{document}

\title{Where Businesses Thrive: Predicting the Impact of the Olympic Games on Local Retailers through Location-based Services Data}

\author{
Petko Georgiev\\
Computer Laboratory \\
University of Cambridge, UK\\
petko.georgiev@cl.cam.ac.uk\\
\And 
Anastasios Noulas \\
Computer Laboratory \\
University of Cambridge, UK\\
anastasios.noulas@cl.cam.ac.uk\\
\And
Cecilia Mascolo\\
Computer Laboratory \\
University of Cambridge, UK\\
cecilia.mascolo@cl.cam.ac.uk\\
}

\maketitle
\begin{abstract}
\begin{quote}
The Olympic Games are an important sporting event with notable consequences for the general economic landscape of the host city. Traditional economic assessments focus on the aggregated impact of the event on the national income, but fail to provide micro-scale insights on why local businesses will benefit from the increased activity during the Games.
In this paper we provide a novel approach to modeling the impact of the Olympic Games on local retailers by analyzing a dataset mined from a large location-based social service, Foursquare. We hypothesize that the spatial positioning of businesses as well as the mobility trends of visitors are primary indicators of whether retailers will rise their popularity during the event. To confirm this we formulate a retail winners prediction task in the context of which we evaluate a set of geographic and mobility metrics. We find that the proximity to stadiums, the diversity of activity in the neighborhood, the nearby area sociability, as well as the probability of customer flows from and to event places such as stadiums and parks are all vital factors. Through supervised learning techniques we demonstrate that the success of businesses hinges on a combination of both geographic and mobility factors. Our results suggest that location-based social networks, where crowdsourced information about the dynamic interaction of users with urban spaces becomes publicly available, present an alternative medium to assess the economic impact of large scale events in a city.
\end{quote}
\end{abstract}

\section{Introduction}
\label{sec:intro}

The Olympic Games are a major international sporting event that involves the large investment of money in providing sporting facilities, transport infrastructure, housing and maintenance. Economic impact assessment reports~\cite{blake2005economic,Lee2005595} have been used as the primary means of evaluating the effects of the Olympic Games on the general economic landscape of a country. Many reports rely on the design of complex financial models, often require resource consuming surveys and aim to assess the overall impact of the event on the national income. However, they rarely provide any insights on how concrete retailers in the host city will be affected. 

The problem of identifying early what businesses will benefit from the increased customer activity during the Games, as well as understanding why, is one with a significant geo-commercial impact for the local economy. 
First, uncovering beforehand the most desirable destinations for users demanding a particular service on the event days could aid location-based advertising. Second, the insights from the analysis of why certain retailers have the potential to expand their market share during a major event can be monetized for the benefits of both services such as Foursquare and retailers. On the one hand, location-based services can act as providers of analytics on local user activity. On the other hand, the businesses in the city can exploit this knowledge to assess the impact of large scale events on their customer flow and improve their future marketing strategy. 

In this work we take advantage of a dataset collected from a popular location-based social service, Foursquare, to address the important challenge: \emph{what are the factors determining whether local businesses will experience a rise in potential customers during the sporting event}? We focus on the most recent London 2012 Olympic Games and analyze user activity through Foursquare check-ins. The check-ins are the location broadcasts publicly shared by users of mobile devices with an installed Foursquare application. Over time, the location-based service has accumulated massive volumes of user-generated content consisting of granular timestamped information regarding users' visits to places such as stadiums, restaurants and shops. The data provided by this new generation of mobile services provides the tools to study the impact of major events on the retail activity of the host city.

\begin{figure*}[!ht]
        \begin{minipage}[b]{0.46\textwidth}
                \centering
                \includegraphics[width=\textwidth]{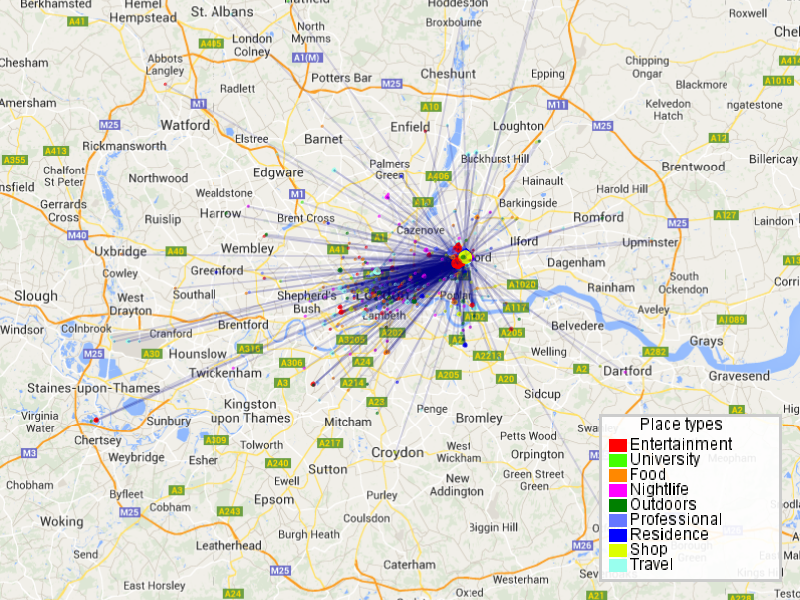}
                \small{(a) Pre-Olympic period}
                \label{fig:transitions_before}
        \end{minipage}%
        \hspace{0.5cm}
        \hfill
        \begin{minipage}[b]{0.46\textwidth}
                \centering
                \includegraphics[width=\textwidth]{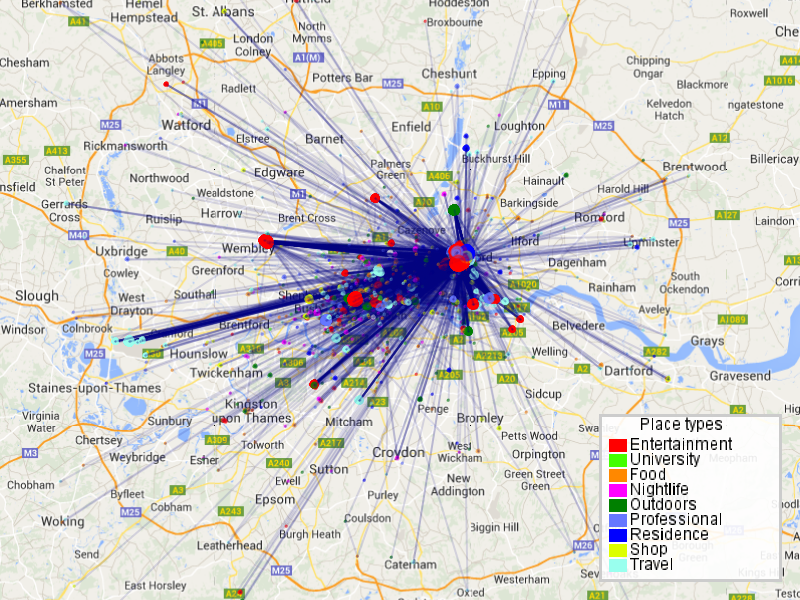}
                \small{(b) Olympic period}
                \label{fig:transitions_during}
        \end{minipage}
        \caption{Transitions towards the Olympic park in Stratford, London in the three-week periods before (4 July - 24 July 2012) and during (25 July - 14 August 2012) the event. Two places are connected if a user subsequently checked in from one venue to another within a day. The size of the points is proportional to the popularity of the place measured in number of check-ins.}
        \label{fig:transitions}
\end{figure*}

Our contributions can be summarized as follows:
\begin{itemize}
\item
We provide an in-depth analysis of the London mobility landscape during the Olympic Games to discover that the major event has a profound impact on the popularity of places measured through Foursquare user check-ins. Our results suggest that the venue popularity rankings from subsequent time periods see their lowest agreement around the Olympic period and at places that are close to where the event itself and live broadcasts are held.
\item
In order to unravel the factors driving the success of local businesses, we formulate a binary classification task. Its aim is to predict, based solely on past check-in data, what food businesses will increase their customers base during the Games. We hypothesize that \emph{the purely spatial advantage of places as well as the mobility trends of visitors can be both highly indicative of the retail winners during the major event}. To test these assumptions, we devise a set of generalizable metrics that exploit geographic and mobility information, and that assess the impact of the major event on 95 local food retailers around the Olympic hot spots.

\item
Through extensive evaluation we discover that information such as the geographic distance from a sports stadium, the diversity of activities in the nearby area, the neighborhood sociability, and the probability of user transitions from or towards entertainment spots all provide powerful signals in the prediction task. We combine the individual features in a supervised learning model to demonstrate that \emph{the success of food businesses depends on both geographic and mobility factors}. We show that a unifying framework can significantly increase the performance of the best individual predictor (from $0.72$ to $0.80$ in the area under the curve (AUC) score). 
\end{itemize}

Our findings show how online location-based social services can be exploited to model the future economic environment of geographic regions of a city during large social and sporting events. In that respect, retail facility owners could identify not only whether their business will be positively affected during a major event, but also make a diagnosis through location-based analytics on the factors that may play a pivotal role in the attraction of increased customer flows in a similar setting.

\section{Dataset Analysis}
\label{sec:analysis}
In this section we describe the dataset that we have collected to study the impact of the Olympics on the changes in user activity during the event. We provide insights that reveal how a major event can dramatically alter the human mobility landscape of the host city which has important implications for the local economy.

\subsection{Dataset Collection}
Foursquare is currently the most popular location-based social networking service with over 45 million users as of January 2014.\footnote{https://foursquare.com/about} The service allows mobile phone users to \emph{check in} at a specific location and share their whereabouts with friends. The Foursquare application gives users the option to link their accounts with other online social services such as Twitter. We have used the Twitter streaming API to crawl the data for Foursquare users who have explicitly shared their check-ins via Twitter. The gathered dataset spans 9 months worth of data during both 2011 and 2012 (Table~\ref{table:dataprops}). 

\begin{table}[htp]
\small
\centering
\begin{tabular}{ |l|r|r|r| }
  \hline
  \textbf{period} & \textbf{\# users} & \textbf{\# venues} & \textbf{\# check-ins} \\
  \hline       
  Dec 2011 - Sep 2012 & $34,202$ & $52,632$ & $578,232$ \\
  Dec 2010 -  Sep 2011 & $41,397$ & $41,701$ & $533,931$ \\
  \hline
\end{tabular}
\caption{Basic statistics of the two Foursquare datasets.} 
\label{table:dataprops}
\end{table}

\begin{figure*}[htp]
        \begin{minipage}[b]{0.23\textwidth}
                \centering
                \includegraphics[width=\textwidth]{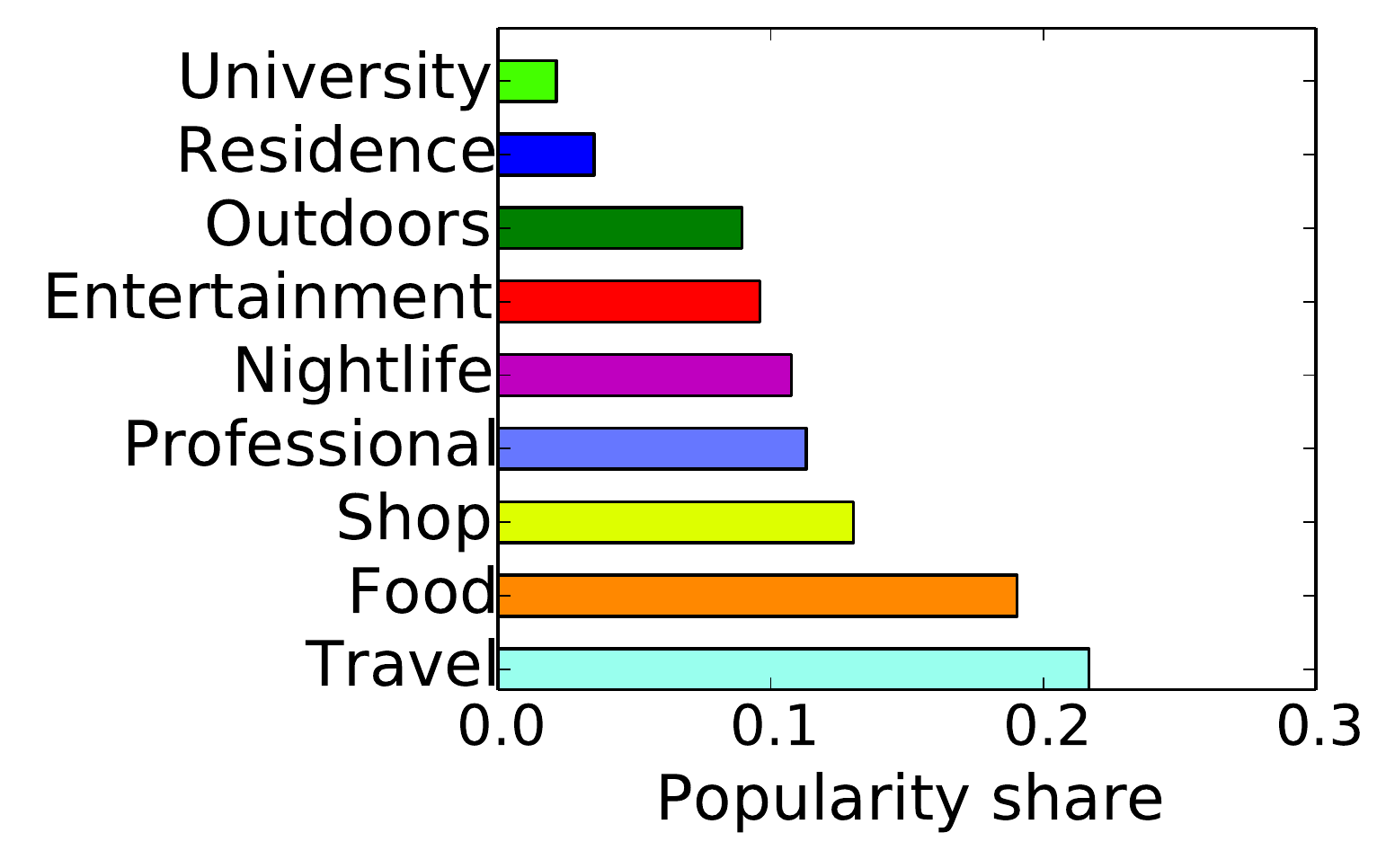}
                \small{(a) Pre-Olympic period, 2012}
                \label{fig:popshare_before}
        \end{minipage}%
        \hspace{0.5cm}
        \hfill
        \begin{minipage}[b]{0.23\textwidth}
                \centering
                \includegraphics[width=\textwidth]{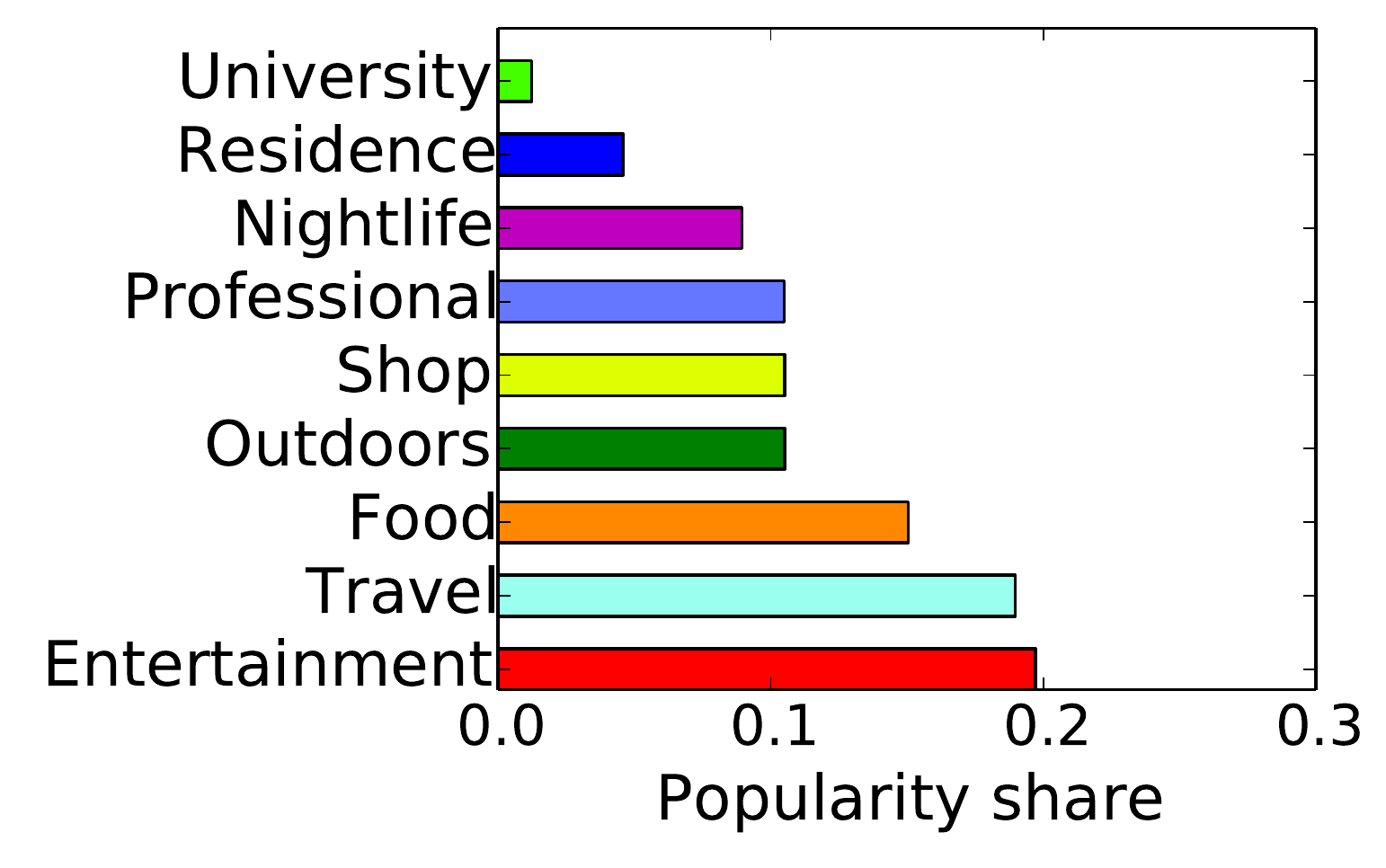}
                \small{(b) Olympic period, 2012}
                \label{fig:popshare_during}
        \end{minipage}
        \begin{minipage}[b]{0.23\textwidth}
                \centering
                \includegraphics[width=\textwidth]{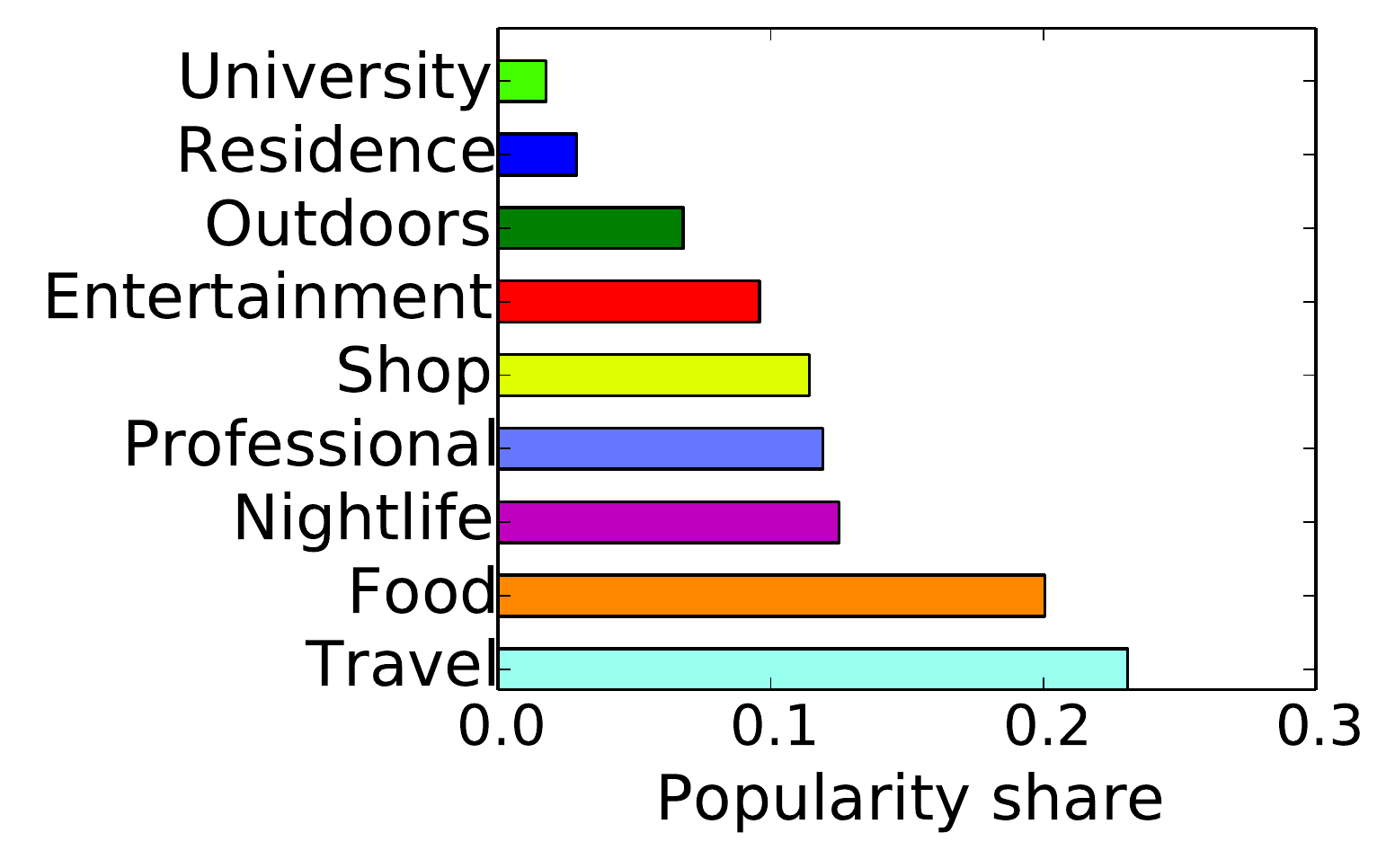}
                \small{(c) Pre-Olympic period, 2011}
                \label{fig:popshare_before2011}
        \end{minipage}%
        \hspace{0.5cm}
        \hfill
        \begin{minipage}[b]{0.23\textwidth}
                \centering
                \includegraphics[width=\textwidth]{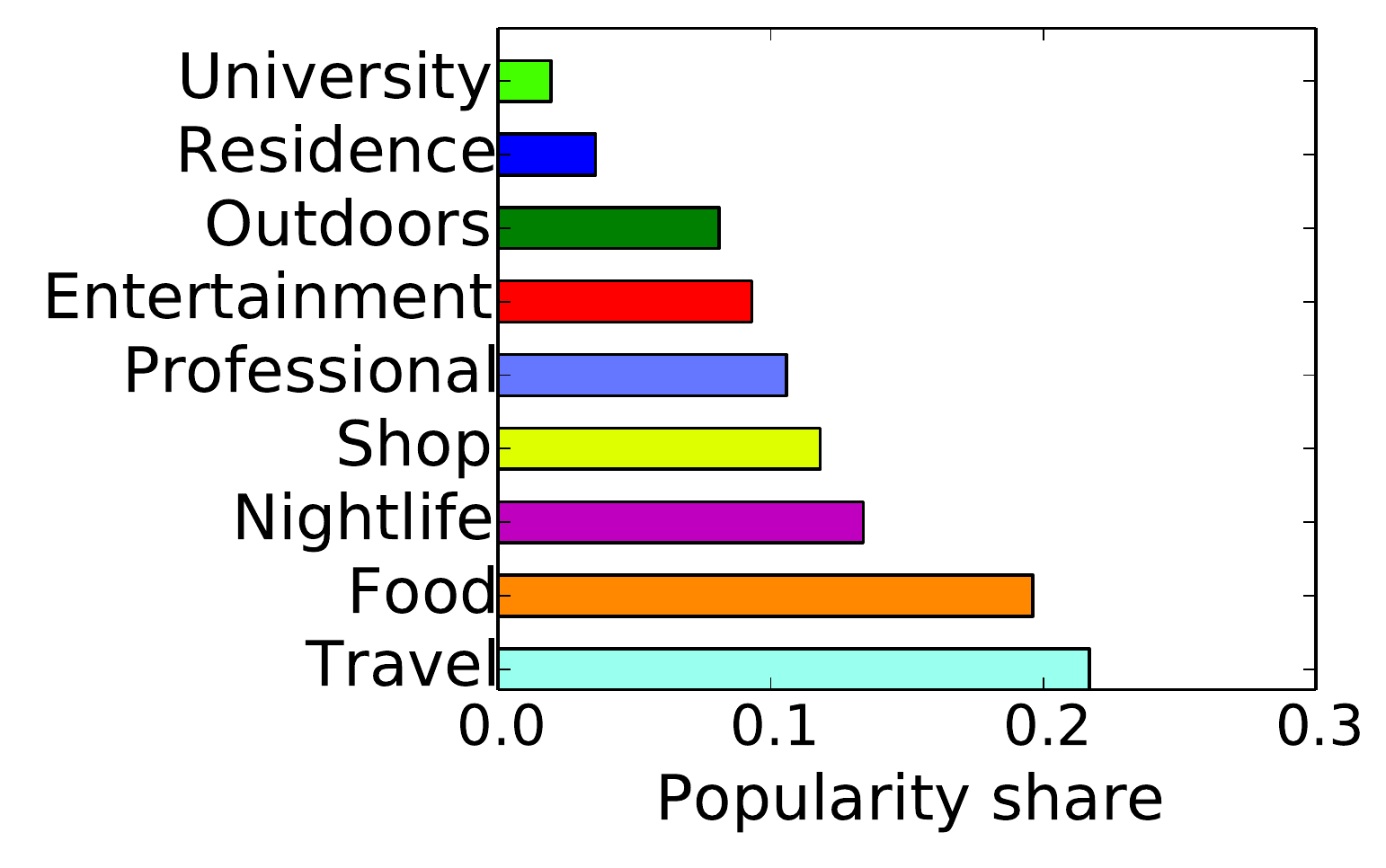}
                \small{(d) Olympic period, 2011}
                \label{fig:popshare_during2011}
        \end{minipage}
        \caption{Relative popularity of the Foursquare places of different categories. The popularity share measures the proportion of check-ins users create at the venues of the corresponding type. We show the scores for the Pre-Olympic (4 July - 24 July) and Olympic periods in 2012 (a-b) as well as their corresponding time spans in 2011 (c-d).}
        \label{fig:popshare}
\end{figure*}

\begin{figure*}[htp]
		\begin{minipage}[b]{0.23\textwidth}
                \centering
                \includegraphics[width=\textwidth]{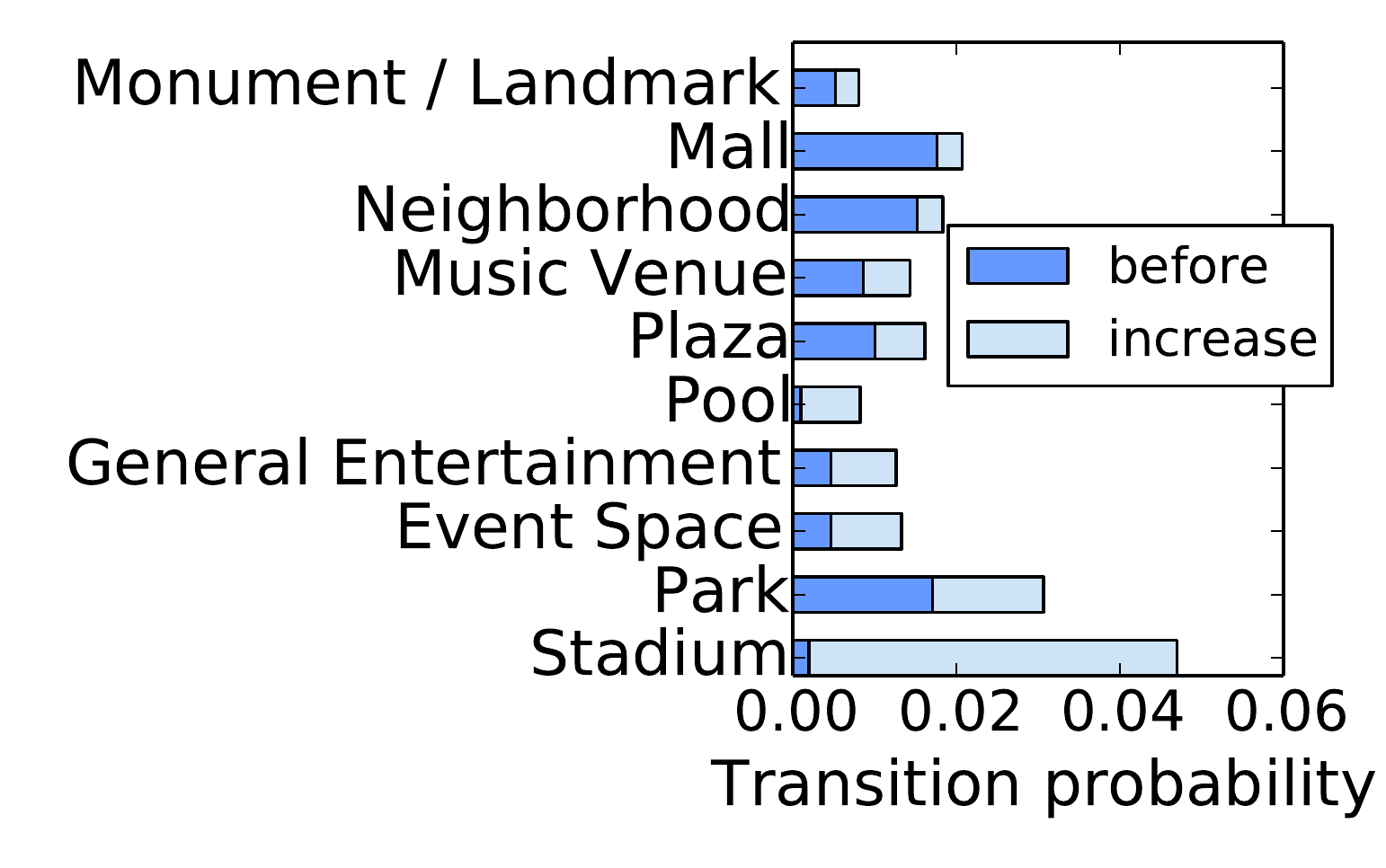}
                \small{(a) \emph{Food}, in-flow increase}
                \label{fig:food_inflow}
        \end{minipage}%
        \hfill
        \begin{minipage}[b]{0.23\textwidth}
                \centering
                \includegraphics[width=\textwidth]{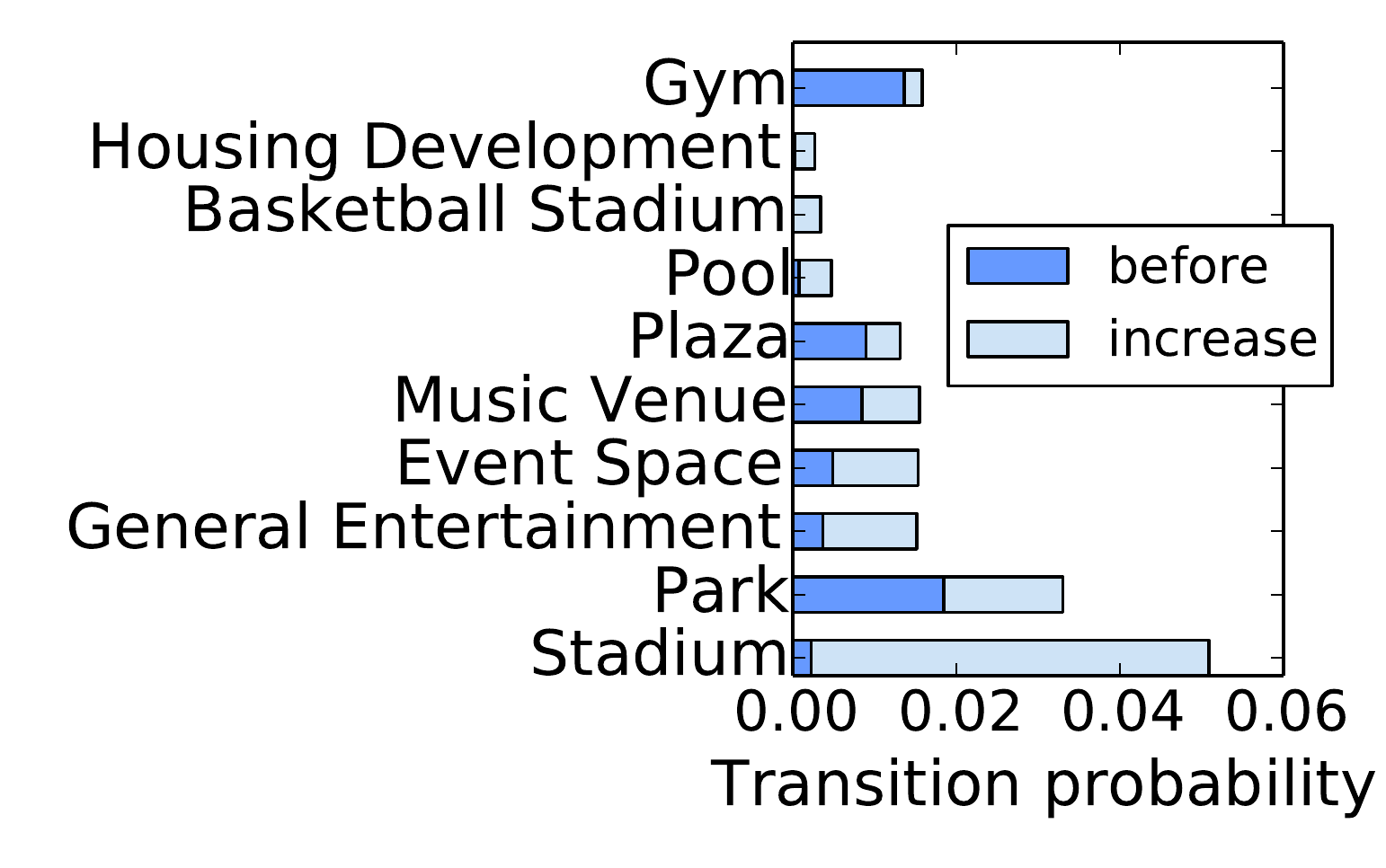}
                \small{(b) \emph{Food}, out-flow increase}
                \label{fig:food_outflow}
        \end{minipage}
        \hfill
        \begin{minipage}[b]{0.23\textwidth}
                \centering
                \includegraphics[width=\textwidth]{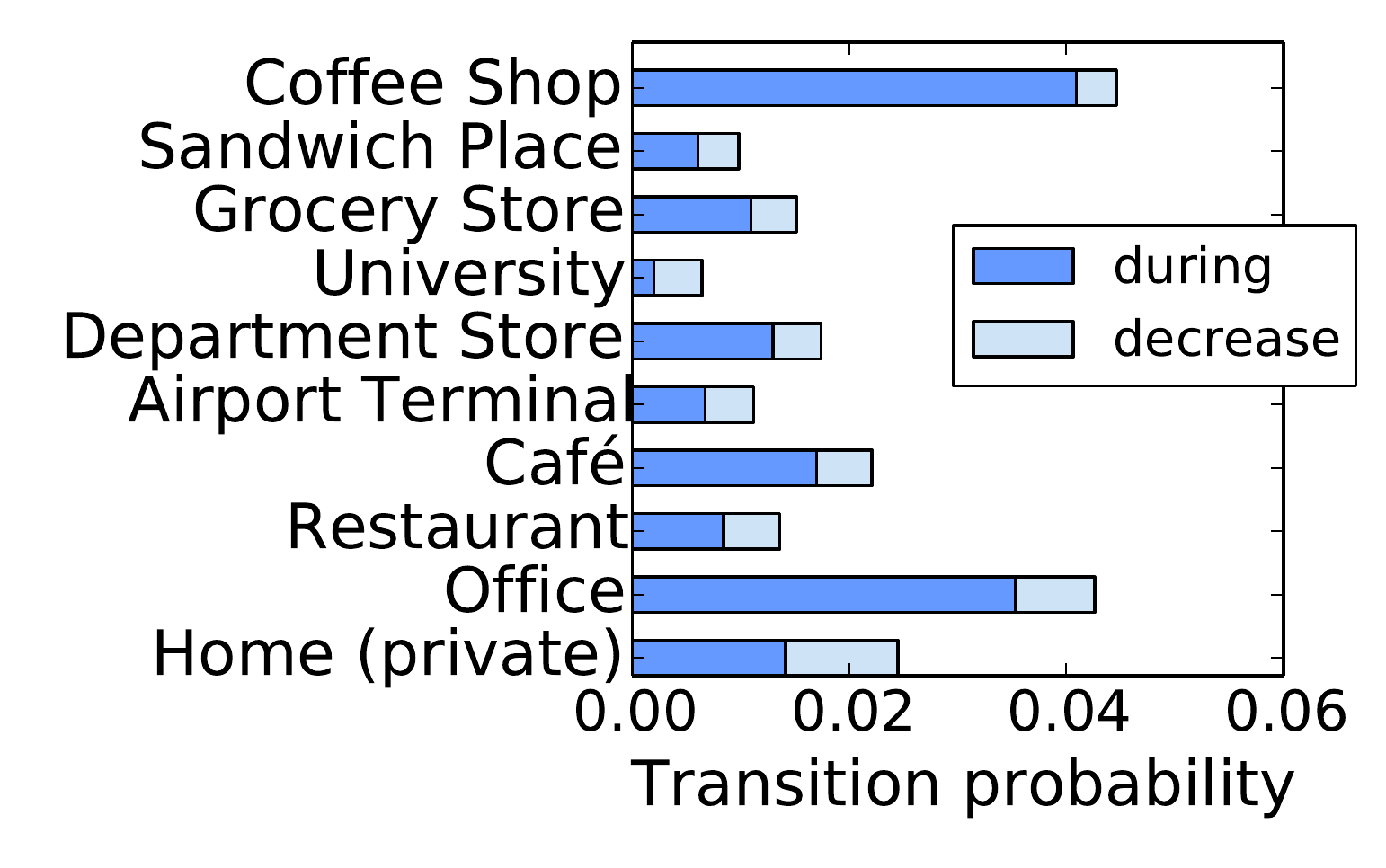}
                \small{(c) \emph{Food}, in-flow decrease}
        \end{minipage}%
        \hfill
        \begin{minipage}[b]{0.23\textwidth}
                \centering
                \includegraphics[width=\textwidth]{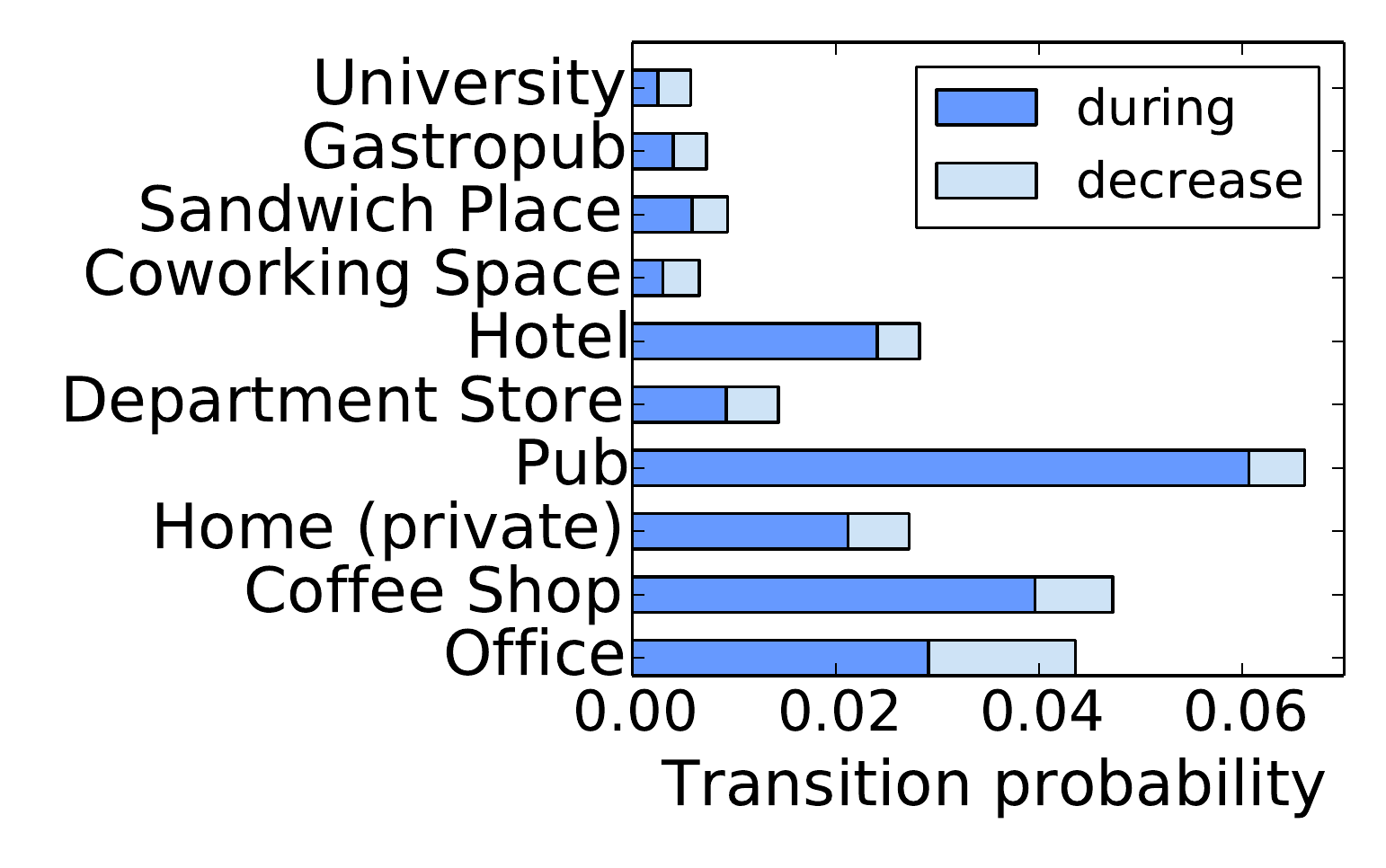}
                \small{(d) \emph{Food}, out-flow decrease}
        \end{minipage}
        \caption{Top increases and decreases in the empirical transition probability from (out-flow) and to (in-flow) \emph{Food} places during the Olympics.}
        \label{fig:flows}
\end{figure*}

The dataset captures the period of the Olympic Games which were held between 27th of July and 12th of August 2012, with these dates marking the days of the opening and closing ceremony respectively. In our analysis we use the data from the previous year to compare the check-in trends around the Olympic period between the two years so that we can alleviate effects related to seasonal biases. From now on, when we refer to the Olympic period, we also include the 2 days before the start and the 2 days after the end in order to obtain a time span of exactly three weeks. 

As Foursquare prohibits unauthorized access to a user's friend lists for privacy reasons, we have obtained friendship information about Foursquare users via their Twitter social network. We consider two users to be friends if they both follow each other on Twitter. While the resulting social graph is not guaranteed to be identical to the Foursquare one, we expect Foursquare users that are connected to each other on Twitter to share some common interests. 

Finally, the tweets contain a URL to the Foursquare website where detailed information about the visited locations is available. We were able to obtain the semantically enriched places corresponding to these locations such as restaurants, stations or shops. Each place belongs to a hierarchy of types, from the more general category such as \emph{Food}, \emph{Entertainment} or \emph{Travel} to the more specific one such as \emph{Coffee Shop}, \emph{Track Stadium} or \emph{Train Station}.

\subsection{The Impact of the Olympics on User Activity}
\label{sec:impact}
The Olympic Games are a major event during which the pulse of the city changes as spectators flood the Olympic park to witness the sporting feats of the top sports people in the world. In Figure \ref{fig:transitions} we demonstrate the formidable rise in the volume of place transitions towards the Olympic venues in Stratford, London where we witness an almost 10-fold increase in the users' movements compared to the Pre-Olympic period. Entertainment and outdoor places such as stadiums, pools and parks are the primary targets of these transitions which is why their popularity is expected to increase during the Games. 

In Figure~\ref{fig:popshare} we substantiate this observation by tracking how the overall popularity of the general place categories changes between the Pre-Olympic and Olympic periods. In the figure we measure the relative popularity of the place types as a proportion of the total number of check-ins created by Foursquare users at the venues during the analyzed period. We observe that \emph{Entertainment} places, which are usually behind \emph{Travel}, \emph{Food}, \emph{Shop}, \emph{Nightlife} and \emph{Professional} places, top the popularity ladder during the Olympics (Figure \ref{fig:popshare}b). In addition, \emph{Outdoors} places experience a jump in the ranking from seventh to fourth most popular category during the Games. We emphasize that these changes are not seasonal as it can be inferred from the London dataset for 2011 where \emph{Travel} and \emph{Food} places remain the top two categories just as they do before the event in 2012. 

Understanding how users move during the Olympics from and to local businesses is an important mobility aspect that we analyze in Figure~\ref{fig:flows}. We display the most pronounced changes in the user flows from and to food places which are the most heavily represented Foursquare category and the main focus of our analysis. We notice that the biggest increases in the empirical probability of transitions happen with respect to the specific place types of the Olympic venues. Transitions from and to \emph{Stadiums}, \emph{General Entertainment} facilities, \emph{Parks}, \emph{Pools} and \emph{Event Spaces} are topping the charts with the biggest rise in popularity. It is notable that these venue types are both sources and targets for the top increases in movements during the Games. This finding is complementary to the general upsurge in entertainment, outdoors and sporting activity discussed in the previous paragraph.

\begin{figure}[htp]
        \centering
        \includegraphics[width=0.46\textwidth]{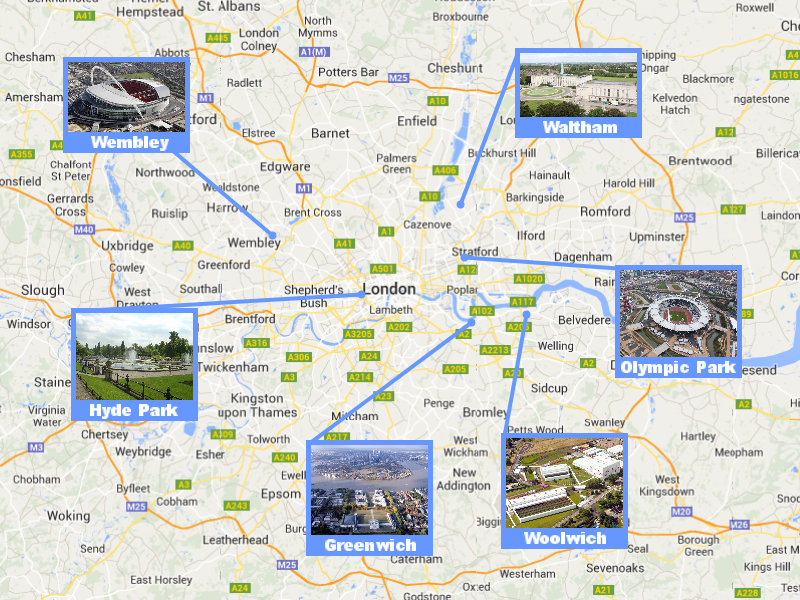}
        \caption{Olympic hot spots: stadiums and live broadcasting sites.}
        \label{fig:hotspots}
\end{figure}

One question that arises from the analysis is whether the Olympic park is the only area in London that experiences increases in user activity due to the Games. If we inspect the data, we notice that there are several major live broadcasting sites which have been captured by the dataset as shown in Figure~\ref{fig:hotspots}. We validate their existence and relation to the Olympics by first looking at the place names which contain the phrases "Live Site 2012" or "Olympic Broadcast Compound", and then by manually searching the documented sites via a web search engine. Not surprisingly the type of these hot spots is specified as \emph{General Entertainment} which, as we have already seen, is one of the top categories that witnesses an increase in its transition probability from and to food places. The most active of the hot spots are the Olympic park in Stratford and Hyde Park in central London, accounting for $55$\% and $29$\% of the user check-ins created at all of the hot spot areas during the Games respectively.

\subsection{Changes in Place Popularity}
\label{sec:pop}
To evaluate the impact of the Olympic Games on the local businesses such as \emph{Food} places, we look at how the popularity levels of these places among Foursquare users change as a function of time. The fluctuations in number of check-ins is used here as a proxy to the different amounts of customers a place such as a restaurant may receive at different times.

\begin{figure}[htp]
        \centering
        \includegraphics[width=0.36\textwidth]{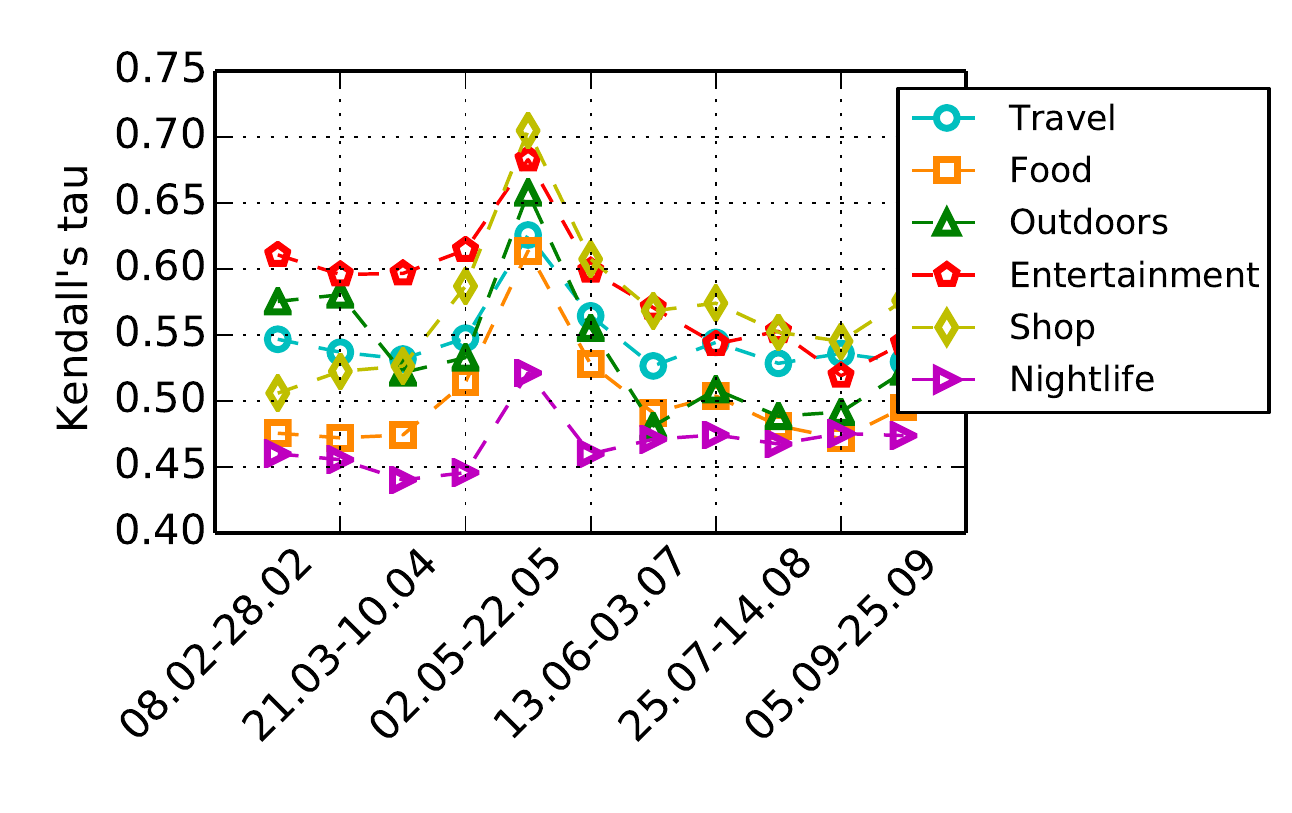}
        \caption{Kendall's tau correlation coefficient for the rankings in popularity of places 
        between two subsequent periods.}
        \label{fig:kendalltau_all}
\end{figure}

We segment the dataset into three-week periods using the Olympic one as a reference (25 July - 14 August 2012)
and rank the places of each general type in descending order of their popularity in each period. We compute the Kendall's tau rank correlation coefficient between the rankings in subsequent periods to quantify how much variation there is in the relative popularity of places between successive time spans. The coefficient values range between -1 and 1 so that a perfect match in the ranking would result in a value of 1.

We observe that the popularity of places in the city remains relatively stable across the different periods with a statistically significant ($p < 0.01$) positive correlation in the rankings between two subsequent periods (Figure \ref{fig:kendalltau_all}). In the face of a major event such as the Olympic Games, it is expected that the areas around the Olympic-related venues will be most affected and it is there that changes in the popularity rankings are most likely. We confirm this by computing the rank correlations between periods as a function of distance to the Olympic hot spots identified in the previous section. 

\begin{figure}[!htp]
        \centering
        \includegraphics[width=0.46\textwidth]{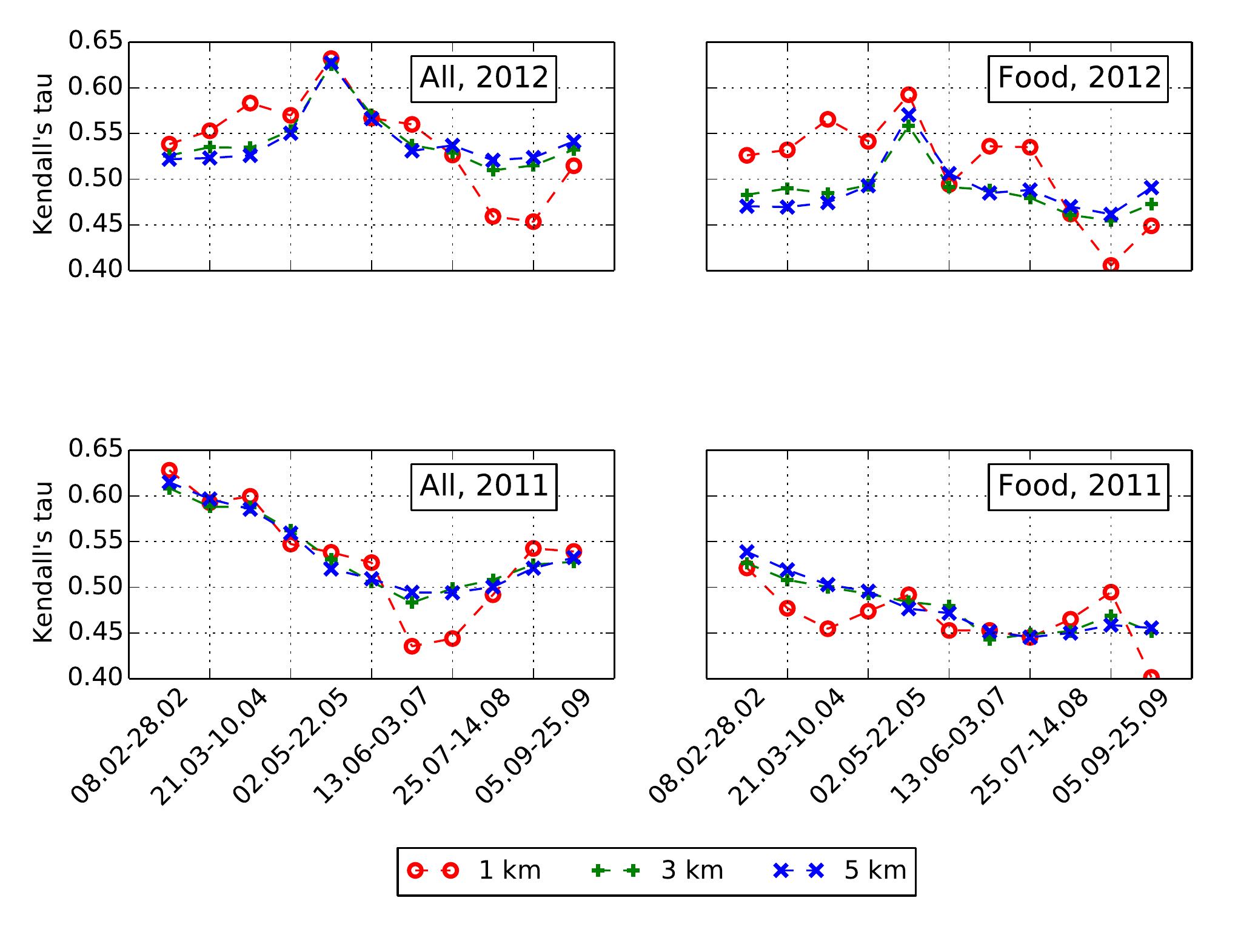}
        \caption{Kendall's tau correlation coefficient as a function of distance to the nearest Olympic live site for the rankings in popularity of all and \emph{Food} places between two subsequent periods.}
        \label{fig:kendalltau_dist}
\end{figure}

Figure~\ref{fig:kendalltau_dist} shows that the most pronounced differences in the popularity rankings are felt in the 1-km vicinity of the Olympic venues. In 2012 the lowest similarity between the neighboring period popularity of all places, including \emph{Food} venues, is felt around the time of the Olympics at close distances. In 2011 we notice that the minimum popularity correlation is observed during periods different from the time span of the Games: early summer for all venues and September for food places. This indicates that the 2012 popularity fluctuations are not due to a seasonal influence. Together these results imply that the Olympic Games disrupt the mobility patterns in the nearby areas where venues are expected to be most affected by the event. Finding what local businesses become more popular during the Olympics, and also understanding the factors that drive this process, are the essential problems we investigate in the following sections.

\section{Predicting Changes in Popularity}
\label{sec:prediction}
Having gained insights about the general check-in pulse of the city during the Olympics, we now motivate a prediction task that aims to forecast what places will increase or decrease their customers during the event compared to the activity trends observed in the previous months.

\subsection{Problem Formulation}
A testimony to the success of a Foursquare venue is its popularity among the service's users which could be used as an approximation to the amount of customer activity the place sees. A larger number of customer visits usually directly translates to an uplift in the total revenue. As a consequence, we consider Foursquare check-ins to be a form of \emph{virtual currency} and a proxy that will enable us to evaluate the impact of the Olympic Games on retail facilities such as restaurants or coffee shops. We elaborate on the implications of this choice of approximating the popularity in the "Discussion and Implications" Section.

To quantify the immediate effects of the Olympic Games on the popularity of retailers, we adopt the \emph{Abnormal Returns} model used in economics \cite{mackinlay1997}. The model measures the impact of a specific event on the value of a firm by observing the market price in the periods before and during the event: the higher the abnormal returns, the higher the impact. In our analysis, we measure the \emph{actual} returns $R_{i\tau}$ for a concrete retailer $i$ as the number of check-ins in the three-week event window $\tau$. We also compute the \emph{expected} returns $E_{i\tau}$ for the event window $\tau$ based on the check-in activity during the three months before the Olympics. $E_{i\tau}$ is effectively the average number of check-ins a place receives for a time window of the same length before the event takes place. We then calculate the abnormal returns $AR_{i\tau}$ as follows:
\begin{equation}
AR_{i\tau} = R_{i\tau} - E_{i\tau}
\end{equation} 

Positive abnormal returns $AR_{i\tau}$ denote a higher than expected popularity of a place, or an increase in the number of customers during the time window $\tau$. Negative values translate to a lower than expected popularity, or a negative event impact. Given the notable geo-commercial importance of finding the local retail winners during the Olympic Games, we aim to predict what places will see positive or negative abnormal returns. The problem reduces to a binary classification task the purpose of which is to separate venues into two groups: the winners that boost their popularity during the Olympics and the adversely affected that experience a loss in absolute terms. Note that we focus on binary prediction rather than regression for two reasons. First in the light of volatile urban dynamics it is a very challenging task to predict accurately the exact number of customers a retail facility will see in a given period. This problem becomes even harder in our case due to the sparsity levels of our check-in dataset.

\subsection{Prediction Space Definition}
Having defined our prediction task and the way to measure the success of local businesses, we proceed with scoping our prediction space to affected retailers. First, we restrict our analysis to places within $1$ km of the Olympic hot spots
because places near the Olympic facilities are most likely to get affected by the event itself. As we have seen in Figure~\ref{fig:kendalltau_dist} (Section "Changes in Popularity"\ref{sec:pop}), the most distinguished changes in the popularity of places occur within close distances of the Olympic-related venues. Second, we focus on \emph{Food} places as they are the most abundant Foursquare category among all in our dataset, constituting more than $23$\% of all documented places in the original dataset. \emph{Food} sites are also the primary source of retail facilities in the Foursquare system, represented by various types of places such us Restaurants, Coffee Shops, Food Trucks, Wineries and many more spots which attendees of nearby Olympic venues could opportunistically visit for refreshments. Overall, there are $88$ Foursquare venue types that fall under the umbrella of the more general category \textit{Food}.\footnote{http://aboutfoursquare.com/foursquare-categories/food/} Finally, we have removed all places with less than $5$ check-ins in the $3$-month period prior to the Olympic Games. Overall we have $95$ instances in the prediction task and a balanced dataset where $48$\% of the instances experience positive abnormal returns.

\section{Prediction Features and Hypotheses}
\label{sec:features}
In the previous section we have motivated and formally defined the problem of predicting what places will be positively or negatively affected during the Olympic Games and have scoped our prediction space. Here we introduce a set of geographic and mobility features to assess the economic potential of the local environment around food retail facilities. The devised features could be generalized in the context of future events and, as a result, their applicability is not constrained to the specific case of the London Olympics evaluated here.

In defining these features we put forward 4 informed hypotheses about the underlying factors contributing to the economic success of retailers during a major event:

\textbf{[\emph{H1}]} \emph{The mobility patterns of users are indicative of whether businesses will increase their number of customers in the event time window.}

\textbf{[\emph{H2}]} \emph{The historical popularity of a place is a weak predictor during a massive event when the activity pulse of the city has dramatically changed.}

\textbf{[\emph{H3}]} \emph{The purely spatial advantage of places is a primary factor in determining the retail winners during the Games.}

\textbf{[\emph{H4}]} \emph{The success of businesses in attracting more potential customers during the Olympics depends on a combination of both geographic and mobility factors.}

\subsection{Notation}
We denote by $P$ the set of places in the host city and with $P(v, r)$ we restrict the set to the ones that are within $r$ meters distance of place $v$. Unless otherwise specified, $r$ is set to $200$ meters as this radius is estimated to be approximately the optimal size of a neighborhood~\cite{urbanNeigh}. For convenience we use $N(v, r) = |P(v, r)|$ to denote the number of places within the area and $N = |P|$ is their total number in the city. The set of specific types such as Coffee Shops or Fast Food Restaurants we annotate with $T$, and by $T_E$ we narrow the set to the types of the Olympic-related venues: \emph{Stadiums}, \emph{General Entertainment}, \emph{Event Spaces}, \emph{Parks}, \emph{Pools}, \emph{Athletics \& Sports} and \emph{Scenic Lookout}. 
To quantify the number of places of a type $t\in T$ within the neighborhood of place $v$ we use $N_t(v, r)$. We introduce a subscript in $t_v$ to denote the specific type of $v$ such as a \emph{Coffee Shop}. The Olympic-related venues and live sites are marked with $v_o$, stadiums are marked with $v_s$, while the restaurants of the major sponsor, McDonald's, are annotated with $v_m$. Last, the set of users visiting a place $p$ are denoted by $U(p)$ and the social network of users is represented by an undirected graph $G(U, E)$ where edges are formed only if the users are both following each other on Twitter.

\subsection{Geographic Features}

The geographic features we introduce assess solely spatial information about places and how their position with respect to others in the neighborhood could contribute positively or negatively to their own popularity profits during the event. 
As a baseline model we use \textbf{\emph{Olympic Distance}} which measures the geographic distance in meters between the target venue $v$ and the closest event-related venue $v_o$: $dist(v, v_o)$. In our case the venue $v_o$ is one of the identified Olympic hot spots. The basic assumption behind the feature is that nearby places will benefit from the increased number of Olympic attendees and spectators. 
To further refine the proximity advantage of event-related venues, we also consider the geographic distance to the nearest stadium $v_s$ and call the feature \textbf{\emph{Stadium Distance}}, $dist(v, v_s)$. Stadiums are the premier gathering point of event attendees for the Olympics just as parks are the event hot spots of summer open-air festivals. The refinement is inspired first by the fact that there are multiple nearby stadiums around the Olympic-related areas of Hyde Park, Olympic Park and Greenwich which can be used for sports training and live broadcasts. Second, the results of the analysis in Section "The Impact of the Olympics on User Activity" \ref{sec:impact} demonstrate that it is the transitions from and towards any type of stadiums that enjoy the biggest increase during the Olympics.

The next two geographic features aim to assess the quality of the neighborhood of a target place. We assume that a neighborhood that is qualitatively better in terms of place type mixture would be a beneficial factor for its venues, especially during a highly active period such as the one of the Olympic Games. We use two established metrics to evaluate the general quality of an area.
The first metric assesses the heterogeneity of a neighborhood in terms of the specific types of places located inside. A more diverse area offers more activities for its visitors which might be particularly relevant in the cases when people stay longer in the area, so as to watch more of the sports games for instance. To measure this place variety, we introduce \textbf{\emph{Nearby Place Entropy}} which has its roots in information theory:
\begin{equation}
-\sum\limits_{t \in T}{\frac{N_{t}(v, r)}{N(v, r)} \times log\frac{N_{t}(v, r)}{N(v, r)}}
\end{equation}
The entropy metric in the location-based context was used by Cranshaw et. al.~\cite{Cranshaw:2010:BGP:1864349.1864380} to evaluate the diversity of unique users visiting a location, while we adopt the notion to model the purely spatial distribution of venues. The higher the entropy, the more bits are required to encode the place type distribution and hence the higher the place diversity is. 

The other feature we employ, \textbf{\emph{Jensen Quality}}, is presented by Jensen~\cite{jensen2006} and evaluates the spatial distribution of places with respect to their ability to attract other venues of certain types, e.g. fast food restaurants next to parks or hotels next to train stations. A neighborhood with higher attractiveness for its target place is expected to be a positive factor for the place popularity and even more so during an active period such as the Olympics. 

The metric uses a utility inter-type coefficient \cite{Jensen:2009:ALR:1617420.1617423} that quantifies the dependency between two types of places in the following manner: 
\begin{equation}
k_{t_p \rightarrow t_v} = \frac{N - N_{t_p}}{N_{t_p}\times N_{t_v}} \sum_{q\in P}{\frac{N_{t_v}(q, r)}{N(q, r) - N_{t_p}(q, r)}}
\end{equation}
Higher scores greater than 1 denote a tendency for the places to attract each other, while lower scores mean that the places tend to repel each other (Table \ref{table:jensen}). The overall quality of a nearby area assesses the desirability of the places around the target venue and is computed as:
\begin{equation}
\sum_{t_p \in T}{k_{t_p \rightarrow t_v} \times (N_{t_p}(v, r) - \overline{N_{t_p}(v, r)})}
\end{equation}
where $\overline{N_{t_p}(v, r)}$ denotes how many places of type $t_p$ are observed on average around places of type $t_v$.

\begin{table}[htp]
\small
\centering
\begin{tabular}{|p{2.7cm}c|lc|}
 \hline
 \textbf{Place type} ($t_v$) & $k_{t_p \rightarrow t_v}$ & \textbf{Place type} ($t_v$) & $k_{t_p \rightarrow t_v}$\\
 \hline
 Wine Shop & 11.620 & Rock Club & 0.040 \\
 Tanning Salon & 10.554 & Mosque & 0.046 \\
 Technology Building & 9.582 & Comedy Club & 0.049 \\
 Car Wash & 5.418 & Dance Studio & 0.055 \\
 Fish Market & 4.217 & Multiplex & 0.057 \\
 Liquor Store & 3.784 & Flower Shop & 0.063 \\
 BBQ Joint & 3.700 & History Museum & 0.064 \\
 Latin Am. Restaurant & 3.363 & Fire Station & 0.074 \\
 Library & 3.342 & Museum & 0.081 \\
 Camera Store & 3.320 & Adm. Building & 0.087 \\
 \hline
\end{tabular}

\caption{Example highest (left) and lowest (right) Jensen attractiveness coefficients for \emph{Fish and Chips Shops}. The values $k_{t_p \rightarrow t_v}$ summarize how frequently places of type $t_v$ are observed around a target place type $t_p$. For instance, \emph{Wine Shops} and \emph{Fish Markets} are commonly found around \emph{Fish and Chips Shops} in London, but this cannot be said for \emph{Rock Clubs} and \emph{Museums}.}
\label{table:jensen}
\end{table}

Another factor we would like to evaluate is the effect of sponsoring venues on other nearby businesses. In the case of the Olympics a major sponsor is McDonald's and a dedicated enormous temporary restaurant is usually built in the Olympic park to serve the event attendees. We compute \textbf{\emph{Sponsor Distance}} as the geographic distance between the target venue $v$ and the closest McDonald's restaurant $v_m$, $dist(v, v_m)$. We expect to understand whether places will benefit from being close to the sponsor which on the Olympic park territory is also the biggest fast food facility and which receives special attention through the Olympic advertising campaigns.

\subsection{Mobility Features}

The mobility aspects we explore attempt to capture how the check-in habits of Foursquare users in the three-week period immediately before the Olympic Games play a role in determining the abnormal returns in popularity in the next period. As a mobility baseline we use place historical \textbf{\emph{Popularity}} which, given the ranking correlations we have seen in Section "Changes in Place Popularity"\ref{sec:pop}, seems a good indicator for the general position of a place in the popularity ladder, but at the same time has a weaker effect during the event.

In Section "The Impact of the Olympics on User Activity"\ref{sec:impact} we have observed that both transitions from and to event-related places such as stadiums, parks and general entertainment facilities experience the sharpest rise in empirical probability among all transitions where food places are involved. That is why we assume that neighborhoods which enjoy a higher amount of transitions from and to entertainment-like places before the Olympics would benefit even more from such movements during the sporting event. 
In order to assess how successful a neighborhood is with respect to its ability to attract flows of users coming from or going to entertainment, sports and outdoor venue types, we introduce the \textbf{\emph{Entertainment Flow}} metric. It computes the mean empirical probability of observing such transitions in the area around a target venue $v$:
\begin{equation}
\frac{1}{N(v, r)} \times \sum_{p \in P(v, r)}{\frac{|\{ \{q, p\}: q \in P \wedge type(q) \in T_{E}\}|}{|\{ \{q, p\}: q \in P\}|}}
\end{equation}
Here $\{q, p\}$ denotes an unordered transition sequence that happens within 24 hours from place $q$ to place $p$ or vice versa.

Many events such as festivals, concerts and sports games are social activities by nature which is why we expect the social motivation for users to attend the Olympic venues to be a strong factor. Research investigating the interplay between social network ties and user movements has confirmed the influence of friends on the general human mobility \cite{Backstrom:2010:FMY:1772690.1772698,Sadilek:2012:FYF:2124295.2124380,friendshipAndMobility}. We design a feature, \textbf{\emph{Social Area}}, that measures the sociability of a neighborhood by counting the pairs of friends that have visited the area in the period before the Olympic Games:
\begin{equation}
|E \cap \{(u_1, u_2): u_1, u_2 \in \cup_{p \in P(v, r)}{U(p)}\}|
\end{equation}
The assumption we make is that the more sociable an area is, the more likely it is to attract friends visiting the nearby Olympic facilities.

\section{Experimental Evaluation}
\label{sec:eval}
In this section we investigate the predictive power of the features defined in the previous section. This allows us to test the introduced hypotheses about the forces driving the economic success of local businesses.

\subsection{Evaluation Methodology and Metrics}

We look at the individual features as unsupervised prediction methods and adopt \emph{precision}, \emph{recall} and Receiver-Operating-Characteristic (ROC) curves as the main tools to analyze prediction performance \cite{Provost:1998:CAA:645527.657469}. ROC curves are non-decreasing plots that summarize how the true positive rate ($\frac{TP}{TP+FN}$) changes as a function of the false positive rate ($\frac{FP}{FP+TN}$). The area under the ROC curve (AUC) is often used as a summary statistic that measures the overall performance of the prediction method. A random classifier would result in a plot that hugs the $y=x$ line where the AUC is $0.50$, while better models would yield curves close to the upper left corner with AUC greater than $0.50$.

To generate the three types of curves, we compute the score each feature gives to a venue in the prediction space and numerically rank the candidates in increasing or decreasing order depending on the directionality of the feature. The distance-based features are ranked in ascending order so that the lowest scores, which are expected to yield the better positive change results, are positioned first in the list. The other features use the descending direction. Given a decision threshold, positive effects are predicted for all candidates with scores lower (or higher) than the threshold. As we vary the threshold we receive different true positive and false positive rates which allows us to build the curves. The \emph{precision} measures the fraction of positive predictions that are correct, while \emph{recall} computes the fraction of positively affected places that are truly predicted. We must note that we are interested in identifying which places will benefit as this could bring up the relevant factors to be taken into account when building advertising campaigns, for instance.

\subsection{Evaluation Results}
\label{sec:ind_feature}

\begin{table}[htp]
\small
\centering
\begin{tabular}{ |l|p{3.6cm}|c| }
  \hline
  \textbf{Feature} & \textbf{Description} & \textbf{AUC} \\
  \hline
  \emph{Random} & Random case baseline & 0.50 \\
  \hline
  \multicolumn{3}{|l|}{\textbf{Geographic}} \\
  \hline
  Olympic Distance & Distance to nearest hot spot & 0.48 \\
  Stadium Distance & Distance to nearest stadium & \textbf{0.72} \\
  Jensen Quality & Nearby area attractiveness & 0.69 \\
  Nearby Place Entropy & Activity diversity in the area & \textbf{0.72} \\
  Sponsor Distance & Distance to McDonald's & 0.68 \\
  \hline
  \multicolumn{3}{|l|}{\textbf{Mobility}} \\
  \hline
  Popularity & Pre-Olympic \# check-ins & 0.56 \\
  Entertainment Flow & Transitions to ent. places & \textbf{0.71} \\
  Social Area & \# friend pairs in the area & \textbf{0.71} \\
  \hline
\end{tabular}
\caption{AUC for the single features used as unsupervised prediction models.}
\label{table:auc}
\end{table}

\begin{figure}[htp]
        \centering
        \includegraphics[width=0.46\textwidth]{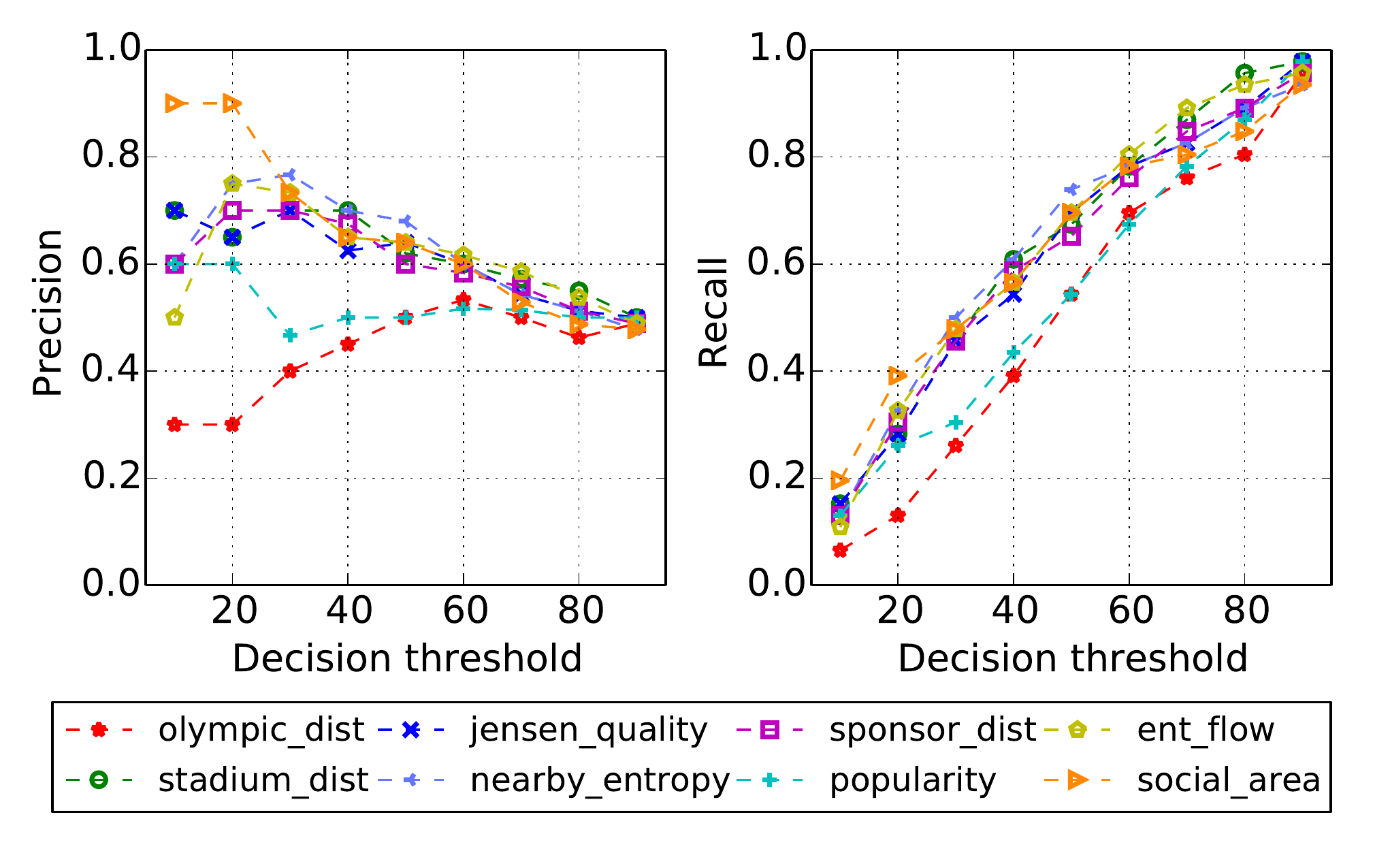}        
        \caption{Precision and recall as a function of the decision threshold for the single features used as unsupervised prediction models.}
		\label{fig:pr}
\end{figure}

\textbf{[\emph{H1}] The Mobility Advantage of Neighborhoods:}
In this part of the analysis we verify our hypothesis that the mobility trends of users visiting the retailers' neighborhoods can reveal whether places will become more popular during the Olympics. We test this by looking at the performance of the introduced mobility features. The \emph{Entertainment Flow} together with the \emph{Social Area} achieve a score of $0.71$ ranking them second highest globally across all features (Table \ref{table:auc}). We recall that we computed the \emph{Entertainment Flow} feature by considering the empirical probability of transitions in the period before the Games towards event-related entertainment venues such as stadiums, parks and pools. The results suggest that the tendency of places in a small neighborhood to be part of a two-hop sequence of visits to recreational venues is a very strong indicator of whether a place would increase its number of customers during a major event such as the Olympics. The venues that already tend to attract visitors from event-related places, be it Olympic sporting facilities or open screenings, are expected to benefit even more when the number of these visitors increases during the event.

We note that the geographic feature \emph{Stadium Distance} is related to this top mobility factor and performs even slightly better with its AUC of $0.72$. This is expected since transitions from and to stadiums are most likely to be observed at short distances. While \emph{Stadium Distance} relies simply on geographic proximity, the mobility feature allows for a wide range of transitions which could occur irrespective of distance. This might account for the better precision and recall results of \emph{Entertainment Flow} at decision thresholds between 20 and 30 items (Figure \ref{fig:pr}).

The \emph{Social Area} feature that models the tendency of the local area to foster social interactions is also a top performer with an AUC equal to $0.71$. In fact, this feature achieves the highest observed precision of $0.90$ at low decision thresholds of up to 20 items. This means that areas that have proven to be historically popular among friends, and especially those with highest sociability scores, are more likely to attract users during the Olympic Games. As organized events are social activities, the nearby areas popular among friends could attract customers attending the sporting shows.

\textbf{[\emph{H2}] The Role of Historical Popularity:}
Here we test our hypothesis that during a massive event such as the Olympic Games, when the activity pulse of particular city regions alters significantly, historical popularity is weak in predicting what food places will benefit from the increased customer activity. We substantiate the claim by the important finding that the \emph{Popularity} baseline performs only marginally better than random for the food places (Table \ref{table:auc}). As we have seen in Section "Changes in Place Popularity"\ref{sec:analysis}, during and immediately after the Olympics the correlation between the popularity rankings of various types of places is lowest at close distances to the event hot spots. A potential interpretation to this behavior is that large scale events can act as game changers on the commercial landscape of a city, and places that have been less popular in the past are provided with a novel opportunity to attract new customer flows.

\textbf{[\emph{H3}] The Spatial Advantage of Retailers:}
In this section we test our hypothesis that the key spatial positioning of businesses is a primary determiner of their success during the event. We confirm this by evaluating the geographic features. A first observation is that the proximity to stadiums is arguably a top factor as already hinted in the discussion about the connection between the \emph{Entertainment Flow} and \emph{Stadium Distance} features. Among the rest of the geographic features, the ones that statically assess the overall neighborhood quality, \emph{Jensen Quality} and \emph{Nearby Place Entropy}, also perform significantly better than random with values for AUC equal to $0.69$ and $0.72$ respectively. We recall that the entropy measured the heterogeneity of an area with respect to its place category mixing. Our results imply that an element of variety in the activities of a neighborhood is a highly positive indicator of whether local food places would boost their customers during the event.

Next, the high AUC score of $0.68$ for \emph{Sponsor Distance} implies that the closer the food venues are to the sponsor, i.e. McDonald's, the more likely they are to benefit. The feature ranks the sponsoring venues first since the measured distance to them is effectively zero. However, the performance is not purely attributed to the sponsor increases in popularity since these venues are only 3 in the prediction dataset. The lower precision results for small decision thresholds of up to $10$ places imply that it is not so much the food venues immediately next to the sponsor that benefit, as there may be an element of competition, but those that are close enough to attract part of the Olympic crowds.

Last, the \emph{Olympic Distance} performs worst and in fact, we cannot claim that it fares better than random. Unlike \emph{Stadium Distance} which can benefit both from the multitude of stadium venues and bigger crowds that these sporting facilities accommodate, \emph{Olympic Distance} is focused on the several live sites around which we build our nearby-venue prediction space. As we are already looking at close distances of no more than 1 km to the Olympic hot spots, further refining the proximity by a few hundred meters does not lead to improvements in performance. 

\textbf{[\emph{H4}] The Interplay of Geographic and Mobility Aspects:}
In this section we test our hypothesis that the success of retailers is dependent on a combination of geographic and mobility factors. We combine the features into a supervised learning model that aims to predict based on the abnormal returns model what places will positively change their popularity and what will not. We assemble a training set of venue feature vectors labeled positively (+1) or negatively (-1) depending on the sign of the abnormal returns $AR_{i\tau}$ and supervise our models to discriminate between the two classes. Our goal is to build a model that achieves a better predictive power than the individual features. 

We compare several algorithms implemented in the WEKA machine learning toolkit \cite{Weka}: Na\"{i}ve Bayes, Random Forests (64 trees, 4 random features each on the full set and 3 random features on the geographic or mobility only) \cite{Breiman01} and Support Vector Machines ($\nu$-SV Classification with probability estimates and $\nu=0.5$)  \cite{Scholkopf:2000:NSV:1139689.1139691}. We exclude the bottom geographic feature, \emph{Olympic Distance}, from the supervised learning task as it does not give performance that is significantly better than the random case. We evaluate the classifiers on the following sets: geographic features only (G), mobility features only (M), and mixed (GM) where the previous two sets are united. Our goal is not only to assess  how the union of all features performs, but also to understand how different types of information sources (geographic vs. mobility) cope with the prediction task.

\begin{table}[!htp]
\small
\centering
\begin{tabular}{ |c|c|c|c|c| }
  \hline
  \textbf{Algorithm} & \textbf{Set} & \textbf{Precision} & \textbf{Recall} & \textbf{AUC} \\
  \hline
  & G & 0.60 & 0.74 & 0.69 \\
  Na\"{i}ve Bayes & M & 0.69 & 0.44 & 0.72* \\
  & GM & 0.74 & 0.63 & 0.72* \\
  \hline
  & G & 0.61 & 0.65 & 0.72* \\
  Random Forest & M & 0.62 & 0.63 & 0.68 \\
  & GM & 0.74 & 0.67 & 0.78* \\
  \hline
  & G & 0.68 & 0.65 & 0.74* \\
  SVM & M & \textbf{0.81} & 0.74 & 0.79* \\
  & GM & 0.71 & \textbf{0.76} & \textbf{0.80*} \\
  \hline
\end{tabular}
\caption{Precision, recall and AUC on the positive items for several supervised learning models on the three different prediction sets. Values at least as high as the AUC of the best feature in the set are marked with an asterisk.}
\label{table:classifiers}
\end{table}

We evaluate the supervised learning models through leave-one-out cross validation which corresponds to an approximately unbiased estimator of the generalization error \cite[p. 260]{hastie}. The metrics we use to compare the classifiers are AUC, \emph{precision} and \emph{recall} computed over the positive samples.
We present our results in Table \ref{table:classifiers}. Although there is some variability in the classifier performance across all metrics, the best results in terms of AUC and recall are achieved when both the geographic and mobility features are taken into account. Random Forests and SVM outperform the best single features, \emph{Stadium Distance} and \emph{Nearby Place Entropy}, in the GM set in terms of AUC with values between $0.78$ and $0.80$ exceeding considerably the single predictors' score of $0.72$. When only geographic or mobility features are used the SVM classifier also succeeds in achieving higher performance than the best feature in the corresponding set. While the SVM algorithm on the mobility set reaches the highest observed precision of $0.81$, it retains lower recall and AUC values than the combined case. The mobility factors offer a good discriminative power as already seen in the evaluation of the individual predictors, but may not be enough to retrieve a larger set of the positively affected businesses which is captured through higher levels of recall. To summarize, these facts together imply that the interactions within the geographic subset, within the mobility subset, and across both sets play a role in determining what food places will improve their popularity and by extension their revenue during the Olympics.

\section{Discussion and Implications}
\label{sec:discussion}

The analysis and evaluation of the influence of geographic and mobility aspects on the popularity changes of food places has revealed interesting insights on the forces driving the increase in the customers base of retail facilities during the Olympics. We acknowledge that the choice of measuring the popularity of places through Foursquare check-ins limits the retail winner prediction space to places where Foursquare users are willing to broadcast their whereabouts. Although the exact popularity figures are not readily available to us, approximating the customer trends through check-ins may prove reasonable in the case of a major event when the upsurge in activity is likely to affect the general population and not only users of the Foursquare service.

All of the features we have designed can be applied either directly or with minor modifications to model the impact of future major events, including upcoming Olympic Games. For instance, the features that appear to be Olympic specific, such as \emph{Stadium Distance} and \emph{Entertainment Flow}, can be trivially altered to measure the distance and user flow to other venues relevant for the domain of other events such as using parks for festivals. While we cannot provide guarantees on the exact performance of the features in future contexts, we expect the general trend of a combination of similar mobility and geographic factors to be most revealing about the event influence on local businesses.

Through extensive evaluation we have revealed that the spatial advantage of places expressed in proximity to stadiums and diversity in nearby activities, as well as neighborhood sociability and historical transitions from and to recreational places are highly effective indicators of whether a food place would increase its popularity during the event. The complexity of the problem of predicting what places will benefit during the Olympics based on historical trends implies the interplay of multiple contributing factors that dynamically interact. This has also been confirmed by our supervised learning approach where the individual signals are fused together. More importantly, we have demonstrated top performance in the AUC summary score when both geographic and mobility aspects are considered in the learning algorithms. This suggests that in combination the features capture non-trivial factor relationships and that location-based services can be employed to predict economic trends of local businesses. This opens a new dimension of modeling possibilities for future major events.

\section{Related Work}
\label{sec:related}
The power of social media for the automated analysis of real-world events has been universally recognized \cite{conf/icwsm/ChakrabartiP11}. The large volumes of timely user-generated content in response to public events such as election campaigns allow the extraction of event insights not easily obtainable via alternative means \cite{DBLP:conf/icwsm/HuJSW12,conf/icwsm/LivneSAA11}. Social media users act as sensors that empower the development of real-time event detection methodologies \cite{weng2011event,Lanagan_Smeaton_2011}. We follow the trends of employing social media for our analysis, and take advantage of location-based services data to automate the impact assessment of a major sports event.

One of our main hypothesis is that geographic factors can play a major role in the popularity of places during highly active seasons such as the Olympic Games. An inspiration is drawn from Jensen's work on quantifying the optimal spatial positioning of retail stores \cite{Jensen:2009:ALR:1617420.1617423,jensen2006}. Jensen as well as Porta et. al. \cite{Porta2009,RePEc:sae:urbstu:v:49:y:2012:i:7:p:1471-1488} demonstrate that pure spatial organization can be indicative of the quality of retail and economic activities in the cities of Lyon (France), Bologna (Italy) and Barcelona (Spain).

Our work is further related to a stream of research on \emph{urban mining} that aims to extract insights from location and mobility data in order to aid urban planning and the development of smart cities. Lathia et al. \cite{Lathia:2012:HIC:2367310.2367318} investigate the interplay between citizen mobility and the well-being of London's census areas through a public transport fare dataset.
Cranshaw et al. and Yuan et al. \cite{conf/icwsm/CranshawSHS12,Yuan:2012:DRD:2339530.2339561}, on the other hand, focus on mobility pattern modeling to characterize the structure of cities with respect to neighborhood dynamics and functional regions. Our approach to extracting mobility features in a neighborhood is related to these works in the sense that we qualitatively assess micro-areas in the city to uncover the underlying factors contributing to venue popularity during the Olympics.

\section{Conclusions}
\label{sec:conclusion}
In this paper we have studied the problem of understanding why certain food businesses will increase their customers during the London Olympic Games in 2012. The considerable economic and commercial benefits of solving the problem with respect to future major events has motivated us to seek the adoption of an alternative source of insights in the face of location-based services where publicly available location data is abundant. We have designed a range of geographic and mobility features that assess the spatial advantage of a venue and have demonstrated that a supervised learning model that combines them has proven particularly effective in the retail winners prediction task. We observe outstanding performance when both the geographic and mobility features are considered. This proves that we have successfully captured core relationships among different factors and that the applicability of our approach extends to future events of a similar magnitude such as the upcoming summer Olympic Games in 2016.

\section{Acknowledgments}
We acknowledge the support of Microsoft Research and EPSRC through grant GALE (EP/K019392). 

{
\small
\bibliographystyle{aaai}
\bibliography{./Bibliography} 
}

\end{document}